\documentclass[conference]{IEEEtran}
\IEEEoverridecommandlockouts

\usepackage{cite}
\usepackage{amsmath}
\usepackage{amssymb}
\usepackage{amsfonts}
\usepackage{graphicx}
\usepackage[dvipsnames]{xcolor}
\usepackage[a4paper, total={184mm,239mm}]{geometry}
\usepackage[english]{babel}
\usepackage[babel=true,final]{microtype}
\usepackage{import}
\usepackage[free-standing-units,per-mode=repeated-symbol]{siunitx}
\usepackage{glossaries}
\usepackage[capitalise,nameinlink,noabbrev]{cleveref}
\usepackage{pgfplots}
\usepackage{pgfplotstable}
\usepackage{tikz}
\usepackage{booktabs}
\usepackage{textcomp}
\usepackage{enumitem}
\usepackage[hyphens]{url}
\usepackage{nth}
\usepackage[version=4]{mhchem}
\usepackage{newtxmath}
\usepackage{multirow}
\usepackage{stfloats}
\usepackage{xspace}
\usepackage{colortbl}
\usepackage[normalem]{ulem}

\usepackage[referable]{threeparttablex}
\renewlist{tablenotes}{enumerate}{1}
\makeatletter
\setlist[tablenotes]{label=\tnote{(\alph*)},ref={(\alph*)},itemsep=\z@,topsep=\z@skip,partopsep=\z@skip,parsep=\z@,itemindent=\z@,labelindent=\tabcolsep,labelsep=.2em,leftmargin=*,align=left,before={\footnotesize}}
\makeatother

\usetikzlibrary{scopes, calc, shapes, arrows, patterns, positioning}
\tikzset{>=latex}

\pgfplotsset{compat=1.18}

\makeatletter
\let\MYcaption\@makecaption
\makeatother
\usepackage[font=footnotesize]{subcaption}
\makeatletter
\let\@makecaption\MYcaption
\makeatother

\makeatletter
\g@addto@macro{\UrlBreaks}{\UrlOrds}
\makeatother

\DeclareSIUnit\bits{bits}
\DeclareSIUnit\bps{bps}
\DeclareSIUnit\Bps{Bps}
\DeclareSIUnit\core{core}
\DeclareSIUnit\tile{tile}
\DeclareSIUnit\request{request}
\DeclareSIUnit\cycle{cycle}
\DeclareSIUnit\erlang{E}
\DeclareSIUnit\flop{FLOP}
\DeclareSIUnit\flops{FLOPS}
\DeclareSIUnit\gate{GE}
\DeclareSIUnit\op{OP}
\DeclareSIUnit\ops{OPS}
\DeclareSIUnit\ipc{IPC}

\definecolor{PULPRed}{HTML}{A8322C}
\definecolor{PULPBlue}{HTML}{1269B0}
\definecolor{PULPGreen}{HTML}{1DB24B}
\definecolor{PULPOrange}{HTML}{F29545}
\definecolor{PULPPurple}{HTML}{910569}
\definecolor{PULPOlive}{HTML}{48592C}
\definecolor{PULPMarine}{HTML}{007996}
\definecolor{PULPGray}{HTML}{ABABAB}
\definecolor{Red}{HTML}{FF0000}

\colorlet{color1}{PULPBlue}
\colorlet{color2}{PULPRed}
\colorlet{color3}{PULPGreen}
\colorlet{color4}{PULPOrange}
\colorlet{color5}{PULPPurple}
\colorlet{color6}{PULPOlive}
\colorlet{color7}{PULPMarine}

\colorlet{colorCore}{PULPRed}
\colorlet{colorMemory}{PULPBlue}
\colorlet{colorInterconnect}{PULPGreen}
\colorlet{colorAccelerator}{PULPOrange}
\colorlet{colorPeripheral}{PULPPurple}
\colorlet{colorAlert}{Red}

\newacronym[longplural={Scratchpad Memories}]{SPM}{SPM}{Scratchpad Memory}
\newacronym{ACE}{ACE}{AXI Coherent Extensions}
\newacronym{AI}{AI}{Artificial Intelligence}
\newacronym{AMBA}{AMBA}{Advanced Microcontroller Bus Architecture}
\newacronym{AMX}{AMX}{Advanced Matrix Extension}
\newacronym{APB}{APB}{Advanced Peripheral Bus}
\newacronym{API}{API}{Application Programming Interface}
\newacronym{ASIC}{ASIC}{Application-Specific Integrated Circuit}
\newacronym{AVX}{AVX}{Advanced Vector Extension}
\newacronym{AXI}{AXI}{Advanced eXtensible Interface}
\newacronym{BLAS}{BLAS}{Basic Linear Algebra Subprograms}
\newacronym{CC}{CC}{Core Complex}
\newacronym{CHI}{CHI}{Coherent Hub Interface}
\newacronym{CMOS}{CMOS}{Complementary Metal-Oxide-Semiconductor}
\newacronym{CNN}{CNN}{Convolutional Neural Network}
\newacronym{CPU}{CPU}{Central Processing Unit}
\newacronym{CSR}{CSR}{Control and State Register}
\newacronym{CTS}{CTS}{Clock Tree Synthesis}
\newacronym{DLP}{DLP}{Data Level Parallelism}
\newacronym{DMA}{DMA}{Direct Memory Access}
\newacronym{DRAM}{DRAM}{Dynamic Random-Access Memory}
\newacronym{DSA}{DSA}{Domain-Specific Accelerator}
\newacronym{DSP}{DSP}{Digital Signal Processing}
\newacronym{DUT}{DUT}{Device Under Test}
\newacronym{ECL}{ECL}{Emitter-Coupled Logic}
\newacronym{FBB}{FBB}{Forward Body-Biasing}
\newacronym{FDSOI}{FD-SOI}{Fully Depleted Silicon on Insulator}
\newacronym{FMA}{FMA}{Fused Multiply-Add}
\newacronym{FPGA}{FPGA}{Field-Programmable Gate Array}
\newacronym{FPU}{FPU}{Floating Point Unit}
\newacronym{GEMM}{GEMM}{General Matrix Multiply}
\newacronym{GPGPU}{GPGPU}{General-Purpose \acrlong{GPU}}
\newacronym{GPU}{GPU}{Graphics Processing Unit}
\newacronym{HDL}{HDL}{Hardware Description Language}
\newacronym{HERO}{HERO}{Heterogeneous Embedded Research Platform}
\newacronym{HPC}{HPC}{High-Performance Computing}
\newacronym{IoT}{IoT}{Internet of Things}
\newacronym{ILP}{ILP}{Instruction Level Parallelism}
\newacronym{IOT}{IoT}{Internet-of-Things}
\newacronym{IPC}{IPC}{Instructions Per Cycle}
\newacronym{IPU}{IPU}{Image Processing Unit}
\newacronym{ISA}{ISA}{Instruction Set Architecture}
\newacronym{LSU}{LSU}{Load/Store Unit}
\newacronym{LVT}{LVT}{low voltage threshold}
\newacronym{MATMUL}{MatMul}{Matrix Multiplication}
\newacronym{GE}{GE}{Gate Equivalents}
\newacronym{MIMD}{MIMD}{multiple instruction, multiple data}
\newacronym{ML}{ML}{Machine Learning}
\newacronym{MMA}{MMA}{Matrix-Multiply Assist}
\newacronym{MME}{MME}{Matrix Multiplication Extension}
\newacronym{MMU}{MMU}{Memory Management Unit}
\newacronym{MRF}{MRF}{Matrix Register File}
\newacronym{MUL}{MUL}{multiplier}
\newacronym{MVL}{MVL}{maximum vector length}
\newacronym{NUMA}{NUMA}{non-uniform memory access}
\newacronym{NOC}{NoC}{Network-on-Chip}
\newacronym{MX}{MX}{Matrix eXtension}
\newacronym{PCIe}{PCIe}{Peripheral Component Interconnect Express}
\newacronym{PC}{PC}{Program Counter}
\newacronym{PE}{PE}{processing element}
\newacronym{PiM}{PiM}{Processing in memory}
\newacronym{PL}{PL}{Programmable Logic}
\newacronym{PMCA}{PMCA}{Programmable Manycore Accelerator}
\newacronym{PnM}{PnM}{Processing near memory}
\newacronym{PNR}{PnR}{Place-and-Route}
\newacronym{PSL}{PSL}{Power Service Layer}
\newacronym{PTE}{PTE}{page-table entry}
\newacronym{PTW}{PTW}{page-table walker}
\newacronym{PULP}{PULP}{Parallel Ultra Low Power}
\newacronym{RAW}{RAW}{read-after-write}
\newacronym{RBB}{RBB}{Reverse Body-Biasing}
\newacronym{ROB}{ROB}{Reorder Buffer}
\newacronym{RTL}{RTL}{Register Transfer Level}
\newacronym{RVT}{RVT}{Regular Voltage Threshold}
\newacronym{RVV}{RVV}{RISC-V Vector}
\newacronym{RoCC}{RoCC}{Rocket Custom Coprocessor Interface}
\newacronym{SCM}{SCM}{Storage Class Memory}
\newacronym{SIMD}{SIMD}{single instruction, multiple data}
\newacronym{SIMT}{SIMT}{single instruction, multiple thread}
\newacronym{SLDU}{SLDU}{Slide Unit}
\newacronym{SLVT}{SLVT}{super-low voltage threshold}
\newacronym{SM}{SM}{Streaming Multiprocessor}
\newacronym{SME}{SME}{Scalable Matrix Extension}
\newacronym{SRAM}{SRAM}{Static Random-Access Memory}
\newacronym{SSE}{SSE}{Streaming SIMD Extension}
\newacronym{SVE}{SVE}{Scalable Vector Extension}
\newacronym{TCDM}{TCDM}{Tightly Coupled Data Memory}
\newacronym{TLP}{TLP}{Thread Level Parallelism}
\newacronym{TxnID}{TxnID}{Transaction ID}
\newacronym{VAC}{VAC}{Vector Access}
\newacronym{VC}{VC}{virtual channel}
\newacronym{VCONV}{VCONV}{Vector Conversion}
\newacronym{VEX}{VEX}{Vector Execute}
\newacronym{VFU}{VFU}{vector functional unit}
\newacronym{VID}{VID}{Vector Instruction Decode}
\newacronym{VIS}{VISSUE}{Vector Instruction Issue}
\newacronym{VLIW}{VLIW}{Very Long Instruction Word}
\newacronym{VLOOP}{VLOOP}{Vector Loop}
\newacronym{VLR}{VLR}{vector length register}
\newacronym{VLSU}{VLSU}{Vector Load/Store Unit}
\newacronym{VNB}{VNB}{Von Neumann Bottleneck}
\newacronym{VRF}{VRF}{Vector Register File}
\newacronym{VPU}{VPU}{Vector Processing Unit}
\newacronym{VT}{VT}{vector thread}
\newacronym{WAR}{WAR}{write-after-read}
\newacronym{WAW}{WAW}{write-after-write}
\newacronym{DCT}{DCT}{discrete cosine transform}
\newacronym{TSV}{TSV}{through-silicon via}
\newacronym{3DIC}{3D-IC}{three-dimensional integrated circuit}
\newacronym{PPA}{PPA}{power, performance, and area}
\newacronym{F2F}{F2F}{face-to-face}
\newacronym{W2W}{W2W}{wafer-to-wafer}
\newacronym{IC}{IC}{integrated circuit}
\newacronym{C4}{C4}{controlled collapse chip connection}
\newacronym{FEOL}{FEOL}{front end of the line}
\newacronym{BEOL}{BEOL}{back end of the line}
\newacronym{PDP}{PDP}{power-delay product}
\newacronym{EDP}{EDP}{energy-delay product}
\newacronym{DRV}{DRV}{design rule violation}
\newacronym{DDR}{DDR}{double data rate}
\newacronym{SDRAM}{SDRAM}{synchronous dynamic random-access memory}
\newacronym{TPU}{TPU}{Tensor-Processing Unit}

\newcommand\ie{i.e.,\xspace}

\newif\ifreviewmode
\newif\ifcamreadymodshow

\ifcamreadymodshow
  \newcommand\cradd[1]{\textcolor{blue}{#1}\PackageWarning{}{#1!}}
  \newcommand\crdel[1]{\sout{\textcolor{red}{#1}}\PackageWarning{}{#1!}}
\else
  \newcommand\cradd[1]{\textcolor{black}{#1}\PackageWarning{}{#1!}}
  \newcommand\crdel[1]{}
\fi

\begin{document}

\title{MX: Enhancing RISC-V’s Vector ISA for Ultra-Low Overhead, Energy-Efficient Matrix Multiplication}
\ifreviewmode
\author{\emph{Hidden for double-blind review purposes.}}
\else
\author{\IEEEauthorblockN{Matteo Perotti\IEEEauthorrefmark{1}\kern-.12em, Yichao Zhang\IEEEauthorrefmark{1}\kern-.12em, Matheus Cavalcante\IEEEauthorrefmark{1}\kern-.12em, Enis Mustafa\IEEEauthorrefmark{1} and Luca Benini\IEEEauthorrefmark{1}\kern-.08em\IEEEauthorrefmark{2}
  \thanks{The first two authors contributed equally to this work.}\thanks{}\thanks{\textcopyright\ 2023 IEEE.  Personal use of this material is permitted.  Permission from IEEE must be obtained for all other uses, in any current or future media, including reprinting/republishing this material for advertising or promotional purposes, creating new collective works, for resale or redistribution to servers or lists, or reuse of any copyrighted component of this work in other works.}}
  \IEEEauthorblockA{\IEEEauthorrefmark{1}ETH Z\"urich, Z\"urich, Switzerland, \IEEEauthorrefmark{2}Universit\`a di Bologna, Bologna, Italy}
  \IEEEauthorblockA{\{mperotti,yiczhang,matheus\}@iis.ee.ethz.ch, emustafa@student.ethz.ch, lbenini@iis.ee.ethz.ch}}
\fi
\maketitle

\begin{abstract}

Dense \gls{MATMUL} is arguably one of the most ubiquitous compute-intensive kernels, spanning linear algebra, DSP, graphics, and machine learning applications. Thus, \gls{MATMUL} optimization is crucial not only in high-performance processors but also in embedded low-power platforms.
Several \glspl{ISA} have recently included matrix extensions to improve \gls{MATMUL} performance and efficiency at the cost of added matrix register files and units. 

In this paper, we propose \gls{MX}, a lightweight approach that builds upon the open-source \gls{RVV} \gls{ISA} to boost \gls{MATMUL} energy efficiency. Instead of adding expensive dedicated hardware, \gls{MX} uses the pre-existing vector register file and functional units to create a hybrid vector/matrix engine at a negligible area cost ($\mathbf{< 3\%}$), which comes from a compact near-FPU tile buffer for higher data reuse, and no clock frequency overhead. We implement \gls{MX} on a compact and highly energy-optimized \gls{RVV} processor and evaluate it in both a Dual- and 64-Core cluster in a 12-nm technology node.
MX boosts the Dual-Core's energy efficiency by 10\% for a double-precision \numproduct{64x64x64} matrix multiplication with the same FPU utilization ($\mathbf{\approx 97\%}$) and by 25\% on the 64-Core cluster for the same benchmark on 32-bit data, with a 56\% performance gain.
\end{abstract}


\IEEEpeerreviewmaketitle

\glsresetall

\section{Introduction}\label{sec:introduction}

The exponential growth of the computational requirements in \gls{ML} and \gls{AI} applications is a major challenge for hardware architects. The rise of application-specific accelerators \cite{PECCERILLO2022102561} and \gls{SIMD} programmable systems \cite{AMIRI202083} demonstrates the need for novel architectures able to cope with the rising computational demand. Furthermore, \gls{AI}/\gls{ML} applications also found their way into edge computing, with benefits such as higher privacy, user personalization, and lower power consumption. However, edge-AI/ML systems have the additional challenge of balancing large computational demands against a very tight power envelope and minimal area footprint.
The quest for energy efficiency and cost (\ie area) minimization is even more pressing today since \gls{ML}/\gls{AI} computation at the edge does not only involve inference but also training in the so-called AI on Edge \cite{9052677}.

\gls{MATMUL} is a cornerstone in \gls{ML} and \gls{AI}, and essential in scientific computing, graphics, and \gls{DSP}.
The importance of \gls{MATMUL} is testified by market-leading companies, such as Google, which developed the first \gls{TPU} in 2015 to accelerate matrix operations~\cite{8358031} and updated it in 2018 with an edge-oriented version achieving \SI{4}{\tera\ops} within a \SI{2}{\watt} power envelope \cite{coral-bmark}.
As with other \glspl{DSA} for specific neural-network tasks \cite{7738524}, the \gls{TPU} is an added resource to which a general-purpose processor offloads the workload (for example, through a PCIe interface). 
This brings an area and power overhead that is not affordable in constrained systems at the edge, especially when they need to compute non-\gls{AI} tasks as well.
Moreover, an excessively specialized accelerator risks becoming useless when the ML/AI algorithm evolves.

Most proprietary \glspl{ISA} offer dedicated matrix extensions, such as Arm's \gls{SME}, Intel \gls{AMX}, and IBM \gls{MMA}. Unluckily, the micro-architectural details of the implementations remain company secrets. 
So far, the RISC-V open-source \gls{ISA} features only a vector extension (\gls{RVV}), even though researchers developed multiple unofficial AI/matrix extensions. Still, they add tightly coupled matrix units \cite{VERMA2022100742} or a new matrix register file \cite{thead-mmul-man} used only during matrix operations, which add area and power consumption. 

\gls{RVV} recently showed to be a valid solution to efficiently accelerate \gls{MATMUL} while keeping a well-known programming model to handle diverse data-parallel workloads, also in the embedded domain \cite{Spatz2022}.
Vector processors execute multiple operations with one instruction, amortizing its fetch/decode cost. Moreover, they feature a \gls{VRF} to buffer the vector elements, decreasing the accesses to memory without changing the computational balance for the architecture~\cite{Kung1986}.
Even if the \gls{VRF} helps decrease the power associated with the memory accesses, it is an additional block at the bottom of the memory hierarchy, one of the key drivers for performance and energy efficiency \cite{billdallyHC}. 
Its size can be way larger than the one of a scalar register file, and it is usually connected to multiple functional units in parallel, which leads to energy-hungry interconnects. 
Hence, the \gls{VRF} access-related energy is usually non-negligible \cite{Perotti2022, Spatz2022}.

With this paper, we present \gls{MX}, a non-intrusive \gls{ISA} extension to \gls{RVV} that creates a general-purpose hybrid matrix/vector architecture with minimal area impact and superior energy efficiency. 
To cut the power consumption, we reduce the expensive accesses to/from the \gls{VRF} by featuring a software-transparent lightweight accumulator close to the processing units. \gls{MX} does not add a matrix unit to the architecture but re-uses the already available processing resources to keep the area and energy overhead at its minimum and exploit the energy efficiency savings that come from the reduced \gls{VRF} accesses.

To validate \gls{MX} across multiple domains, we add \gls{MX} to a constrained embedded Dual-Core cluster built upon the open-source energy-optimized \gls{RVV}-based Spatz~\cite{Spatz2022} vector processor and to a scaled-up MemPool architecture~\cite{MemPool2023_journal} with 64 Spatz processors and implement both systems in a competitive 12-nm technology. 
We provide a quantitative justification of the energy savings and a detailed \gls{PPA} analysis on matrix multiplications on different data precisions, finding that our matrix extension can boost not only energy efficiency but also performance.

With this paper, we present the following contributions:

\begin{itemize}
    \item We define \gls{MX}, a lightweight and non-intrusive \gls{ISA} extension based on \gls{RVV} 1.0 aimed at supporting memory and computational operations directly on matrices. \gls{MX} reduces the power consumption of the architecture with similar or better performance by introducing a near-\gls{FPU} tile buffer, a per-vector-element broadcast system, and minimal modifications to the \gls{VLSU}. 
    \item We provide a theoretical justification of the benefits that the \gls{ISA} has on the power consumption when executing a matrix multiplication kernel, effectively reducing the expensive \gls{VRF} accesses.
    \item We implement \gls{MX} on a constrained Dual-Core and a complex 64-core clusters based on the energy-efficient \gls{RVV} vector processor Spatz, and characterize \gls{MX}'s impact on performance and \gls{PPA} metrics in a 12-nm technology. For less than 3\% area overhead, we get a maximum of 56\% and 25\% performance and energy efficiency gains, respectively.
\end{itemize}

\section{Analysis}\label{sec:analysis}

In the following, we discuss the tiling of a \gls{GEMM} problem through a multi-level memory hierarchy. When C is a zero matrix, \gls{GEMM} becomes a \gls{MATMUL}.

\begin{equation}
    D_{M \times N} = A_{M \times K} \cdot B_{K \times N} + C_{M \times N}
\end{equation}

\begin{figure}
    \centering
    \includegraphics[width=\columnwidth]{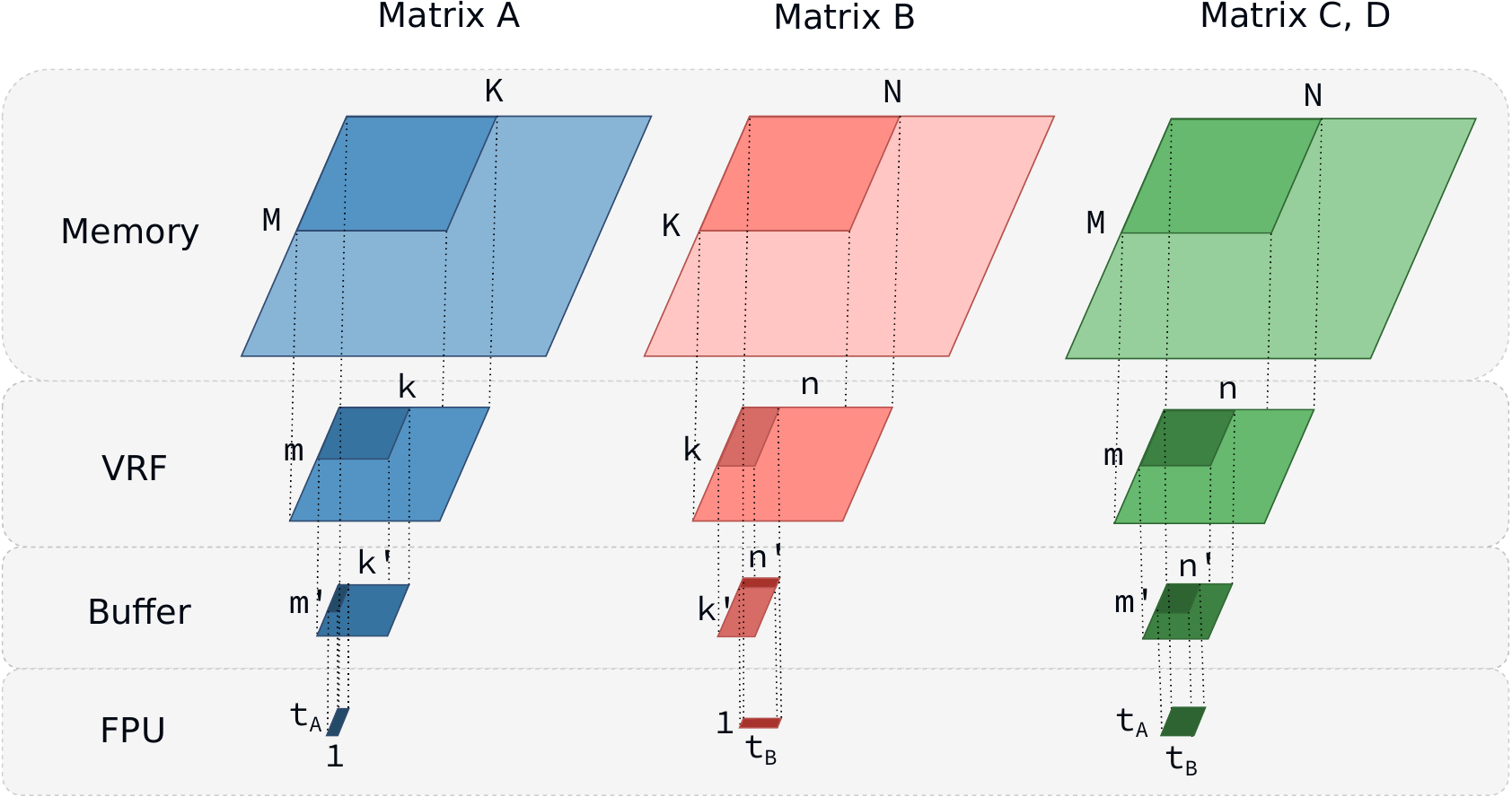}
    \caption{The tiling problem over a memory hierarchy composed of three levels, ending with the processing elements (\glspl{FPU}).}
    \label{fig:mxu-theory}
\end{figure}

For convenience, let us consider a memory hierarchy composed of a memory, a \gls{VRF}, and a near-\gls{FPU} buffer. The memory connects to the \gls{VRF}, which is connected to a buffer that feeds the \glspl{FPU}, as reported in \Cref{fig:mxu-theory}. 
The following analysis can be easily extended to memory hierarchies with a different number of levels.

\subsection{The tiling problem}
The lower level of the hierarchy is usually not large enough to keep the input and output matrices all at once. Therefore, the matrices are divided into chunks (tiles), and the hardware works on one output tile at a time, and the outer product algorithm is often used to maximize parallelism.
The number of elements transferred between two consecutive levels of the hierarchy impacts both performance and power consumption and depends on how the matrices are tiled. Usually, the number of transfers is partially encoded in the \textit{arithmetic intensity}, \ie the total number of operations divided by the total number of Bytes transferred between the memory and the core.

In the following, we provide equations to fine-grain count how many memory accesses happen between each pair of consecutive levels of the hierarchy. 
Each equation contains four terms, which correspond to 1) the elements of matrix A, 2) the elements of matrix B, 3) the elements of matrix C (or D) from the upper level to the lower one (load/fetch), and 4) the elements of the matrix D from the lower level back to the upper one (store/write-back).

In the most generic scenario, without buffering the output tile in the \gls{VRF} for more than $k$ updates, the number of elements moved between the memory and the \gls{VRF} is:
\begin{equation}
\label{eq:mem-vrf}
  \#Elm^{MEM}_{VRF} = \frac{N}{n}MK + \frac{M}{m}NK + \frac{K}{k} MN + \frac{K}{k} MN
\end{equation}
Where the A, B, and D (C) matrices stored in memory have sizes $MK, NK, MN$, and we tile the problem between the memory and the \gls{VRF} with tiles of size $mk, nk, mn$.

For each matrix tile, the number of elements exchanged between the \gls{VRF} and the buffer is:
\begin{equation}
\label{eq:vrf-buf}
  \#Elm^{VRF}_{BUF} = \frac{n}{n'}mk + \frac{m}{m'}nk + \frac{k}{k'} mn + \frac{k}{k'} mn
\end{equation}
Where the tiles stored in the \gls{VRF} have sizes $mk, nk, mn$, and we sub-tile the problem between the \gls{VRF} and the buffer with sub-tiles of size $m'k', n'k', m'n'$.

For each matrix sub-tile, the number of elements exchanged between the buffer and the \glspl{FPU} is:
\begin{equation}
\label{eq:buf-fpu}
  \#Elm^{BUF}_{FPU} = \frac{n'}{t_B}m'k' + \frac{m'}{t_A}n'k' + k'm'n' + k'm'n'
\end{equation}
Where the sub-tiles stored in the buffer have sizes $m'k', n'k', m'n'$, and we access $t_A$ and $t_B$ elements from tiles A and B, respectively.

\subsection{Total number of transfers} To get the total number of transfers between each pair of hierarchy levels, we need to take into account how many output tiles and sub-tiles we calculate throughout the program. 

\paragraph{Memory and VRF} In the most generic case, we load the C (first iteration) and D (from the second iteration on) tiles from memory before consuming the input $mk, nk$-sized tiles, and we store the D tile back to memory after $k$ updates. Without inter-k-tile buffering in the \gls{VRF}, we load $(\frac{K}{k})\frac{M}{m}\frac{N}{n}$ output tiles with size $mn$ from the memory to the \gls{VRF}, and we store back the same amount. If we buffer the output tiles until they are completely calculated over the whole K dimension, the formula simplifies to $\frac{M}{m}\frac{N}{n}$. Instead, we load a total of $\frac{K}{k}\frac{M}{m}\frac{N}{n}$ input A and B tiles, with sizes $mk, nk$.

\paragraph{VRF and buffer}The $m'n'$ output sub-tiles of a tile are fetched for $(\frac{k}{k'})\frac{m}{m'}\frac{n}{n'}$ times from the \gls{VRF} before consuming the input $m'k', n'k'$-sized sub-tiles in the buffer, and written-back to the \gls{VRF} for the same number of times if there is no inter-k-tile buffering in the buffer. Instead, the input A and B sub-tiles, with sizes $m'k', n'k'$, are loaded $\frac{K}{k}\frac{M}{m}\frac{N}{n}$ times.
If we keep into account how many times each tile is loaded/stored from/to memory, these formulas become $(\frac{K}{k})(\frac{k}{k'})\frac{M}{m'}\frac{N}{n'}$ for the sub-tiles fetch, the same amount for the sub-tiles writes-back, and $\frac{K}{k'}\frac{M}{m'}\frac{N}{n'}$ for the A and B sub-tiles fetch.

\crdel{In \Cref{tab:base-eq}, we summarize all the transfers throughout the hierarchy.}
\cradd{We summarize all the transfers across the hierarchy in \Cref{tab:base-eq}.}

\begin{table}[t]
  \caption{Number of accesses between consecutive levels of the memory hierarchy.}
  \label{tab:base-eq}
  \centering
  \begin{threeparttable}
  \resizebox{\columnwidth}{!}{\begin{tabular}{llrrrr}
  \toprule
  \textbf{Ref.} & \textbf{Metric} & \textbf{A ($\downarrow$)} & \textbf{B ($\downarrow$)} & \textbf{C, D ($\downarrow$)} & \textbf{D ($\uparrow$)} \\
  \midrule
  1) & $\mathrm{\#Elm^{MEM}_{VRF}}$ & $\mathrm{\frac{N}{n}MK}$ & $\mathrm{\frac{M}{m}NK}$ & $\mathrm{\frac{K}{k} MN}$ & $\mathrm{\frac{K}{k} MN}$ \\
  2) & $\mathrm{\#Elm^{VRF}_{BUF}}$ & $\mathrm{\frac{N}{n'}MK}$ & $\mathrm{\frac{M}{m'}NK}$ & $\mathrm{\frac{k}{k'} \frac{K}{k} MN}$ & $\mathrm{\frac{k}{k'} \frac{K}{k} MN}$ \\
  3) & $\mathrm{\#Elm^{BUF}_{FPU}}$ & $\mathrm{\frac{N}{t_B}MK}$ & $\mathrm{\frac{M}{t_A}NK}$ & $\mathrm{k'\frac{k}{k'}\frac{K}{k}MN}$ & $\mathrm{k'\frac{k}{k'}\frac{K}{k}MN}$ \\
  \bottomrule \\
  \end{tabular}}
    \begin{tablenotes}
      \item\label{tnote:arr2} $\downarrow/\uparrow$ indicate transfers to a lower/higher level of the memory.
    \end{tablenotes}
  \end{threeparttable}
\end{table}

\subsection{Optimizations}
\paragraph{Inter-k-buffering} If the output tile (sub-tile) is buffered in the \gls{VRF} (buffer) until the whole K (k) dimension is traversed and the whole output tile (sub-tile) is ready, we can simplify the equations above.
If the buffering happens in the \gls{VRF}, $\frac{K}{k} = 1$ in \Cref{tab:base-eq} Ref. 1), while if it happens in the buffer until the whole $K$ dimension, $\frac{K}{k}\frac{k}{k'} = 1$, in \Cref{tab:base-eq} Ref. 2) (if the buffering only happens until the whole $k$ dimension is traversed, $\frac{k}{k'} = 1$). 

Inter-k-buffering is ultimately limited by the size of the lower memory level, which should be able to host the whole output tile (sub-tile) for the whole computation on the $K$ ($k$) dimension. Therefore, keeping the output tile in the buffer for the whole $K$ dimension requires that $m = m', n = n'$. On the other hand, relaxing this constraint, e.g. $m = m', n = B \times n'$, allows for fewer overall transfers between the memory and the \gls{VRF}. In this case, the inter-k-buffering can be done only between the memory and the \gls{VRF}.

\paragraph{C-tile reset} When \texttt{C} is a zero matrix, it's possible to avoid loading it from memory and initialize the \gls{VRF} with zeroes or reset the buffer. If we use inter-k-buffering and the zero initialization is applied to the \gls{VRF} and the buffer, the third term of the equations related to the load/fetch of matrix C, D, becomes zero in \Cref{tab:base-eq} Ref. 1) and 2), respectively.

\section{MX Implementation}\label{sec:isa-ext}
\subsection{ISA Extension}
\begin{figure}
    \centering
    \includegraphics[scale=0.5]{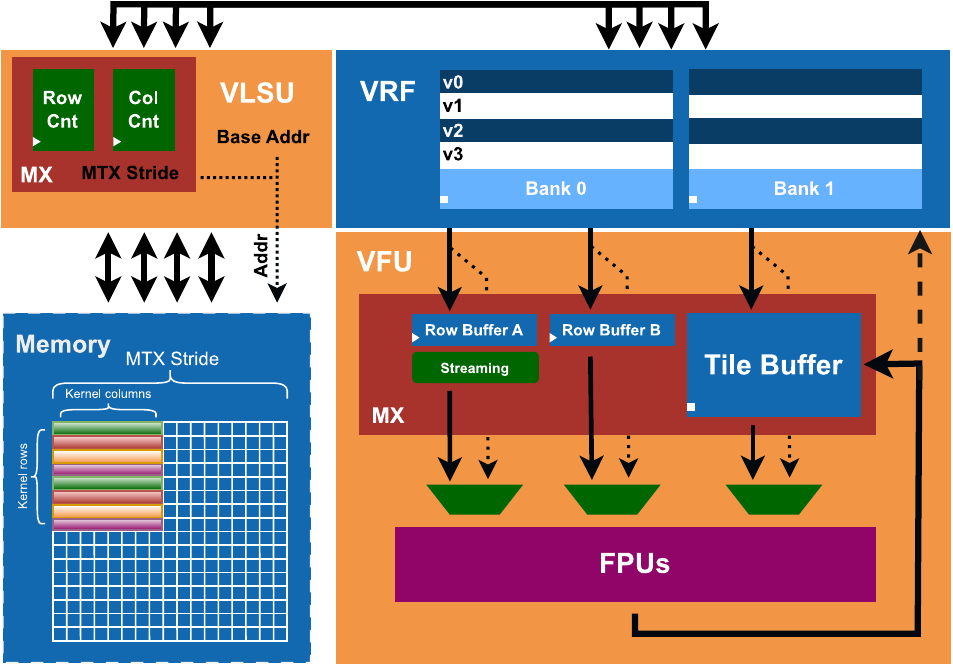}
    \caption{Spatz's \gls{VLSU}, \gls{VRF}, and \gls{VFU} with \gls{MX} architectural schematic.}
    \label{fig:mxu-arch}
\end{figure}

We implement \gls{MX} in Spatz~\cite{Spatz2022}, an open-source, \gls{RVV}-based, highly-optimized compact vector processor, targeting minimal area and power overhead. Thus, we do not add any dedicated matrix units or software-addressable registers, as shown in~\Cref{fig:mxu-arch}.
\gls{MX} adds three configure instructions (\crdel{\texttt{msettilem, msettilen, msettilek}}\cradd{\texttt{msettile[m,n,k]}}), three memory instructions (\crdel{\texttt{mld.a, mld.b, mst.c}}\cradd{\texttt{mld.[a,b]}, \texttt{mst.c}}), and two computation instructions (\crdel{\texttt{mxmacc}, \texttt{mxfmacc}}\cradd{\texttt{mx[f]macc}}). 
The first three instructions set up the sub-tile sizes $m', n', k'$ on which the matrix instructions \crdel{will }operate ($m'k' = vl$, $m'n' \le vl$, and $vl$ is the vector length in elements)\crdel{, writing \gls{MX}-specific \glspl{CSR}}.
We enhance the \gls{VLSU} to enable matrix load and store operations, which are composed of multiple unit- and non-unit-stride memory operations already supported by Spatz.
The \crdel{second set of three instructions is introduced for loading a tile}\cradd{new memory instructions are introduced to load tiles} from matrix \textit{A} and\crdel{ matrix} \textit{B}\crdel{, and for storing} \cradd{and to store} the computed tile back \crdel{into the output matrix}\cradd{to memory}\crdel{.}\cradd{, while the two computational instructions perform a \gls{MATMUL} between the two $m'k'$ and $k'n'$ sub-tiles, storing the resulting $m'n'$ sub-tile in the \gls{VRF}. }Close to the \glspl{FPU}, we introduce a tiny broadcast block consisting only of a register and some multiplexers to broadcast single elements from the \textit{A} tile across multiple elements of tile \textit{B}.
This block increases the data reuse of tile \textit{A} at a minimal cost.
Finally, we implement a latch-based result tile buffer in the \gls{VFU} to reduce energy consumption by minimizing \gls{VRF} accesses by intermediate result accumulation, limiting the buffer size to \(\frac{1}{8}\) of the \gls{VRF} size (i.e., $BUF_{size} = 256 B$) to maintain low energy and area overhead.
Since Spatz's \gls{VLSU} has four parallel memory ports and the buffer is constrained in size, $m', n', k' \in \{4, 8\}$.

\subsection{MX Benefits}
\Cref{tab:comp-acc-en} summarizes the number of elements transferred between consecutive memory hierarchies for a baseline vector-only \gls{MATMUL} and a \cradd{\gls{MX}-ready} \gls{MATMUL}\crdel{ executed with \gls{MX}}.
The baseline employs a traditional scalar-vector algorithm to load \textit{m} scalar elements from the input matrix \textit{A} and an \textit{n}-long vector from matrix \textit{B}.
The \gls{MX}-ready configuration loads A tiles with size $m'k'$ and \textit{B} tiles with size $n'k'$.
While the \gls{MX} algorithm does not further sub-tile the tiles on $m$ or $k$ ($\textit{m'} = \textit{m}$ and $\textit{k'} = \textit{k}$), it sub-tiles along $\textit{n}$ such that $\textit{n} = B \times \textit{n'}$, where $B \in \{2, 4\}$.
In the following, we highlight the benefits brought by the \gls{MX} algorithm.

\begin{table}[t]
  \caption{Data Transfers: MX-ready vs Baseline.}
  \label{tab:comp-acc-en}
  \centering
  \begin{threeparttable}
      \resizebox{\columnwidth}{!}{\begin{tabular}{llrrrr}
      \toprule
      \textbf{Config} & \textbf{Metric} & \textbf{A ($\downarrow$)} & \textbf{B ($\downarrow$)} & \textbf{C, D ($\downarrow$)} & \textbf{D ($\uparrow$)} \\
      \midrule
      Baseline$^\mathrm{\ref{tnote:xrf}}$ & $\mathrm{\#Elm^{MEM}_{VRF}}$ & $\mathrm{\frac{N}{n}MK}$ & $\mathrm{\frac{M}{m}NK}$ & $\mathrm{0}$ & $\mathrm{MN}$ \\
      Baseline$^\mathrm{\ref{tnote:xrf}}$ & $\mathrm{\#Elm^{VRF}_{FPU}}$ & $\mathrm{\frac{N}{F}MK}$ & $\mathrm{MNK}$ & $\mathrm{KMN}$ & $\mathrm{KMN}$ \\
      \midrule 
      MX & $\mathrm{\#Elm^{MEM}_{VRF}}$ & $\mathrm{\frac{N}{B \times n'}MK}$ & $\mathrm{\frac{M}{m'}NK}$ & $\mathrm{0}$ & $\mathrm{MN}$ \\
      MX & $\mathrm{\#Elm^{VRF}_{BUF}}$ & $\mathrm{\frac{N}{n'}MK}$ & $\mathrm{\frac{M}{m'}NK}$ & $\mathrm{\frac{K}{k'} MN}$ & $\mathrm{\frac{K}{k'} MN}$ \\
      MX & $\mathrm{\#Elm^{BUF}_{FPU}}$ & $\mathrm{\frac{N}{F}MK}$ & $\mathrm{\frac{M}{F}NK}$ & $\mathrm{KMN}$ & $\mathrm{KMN}$ \\
      \bottomrule \\
    \end{tabular}}
    \begin{tablenotes}
      \item\label{tnote:xrf} Elements from \textbf{A} are loaded/fetched to/from the scalar register file;
      \item\label{tnote:fpu} \textit{F} represents the number of FPUs;
      \item\label{tnote:arr1} $\downarrow/\uparrow$ indicate transfers to a lower/higher level of the memory.
    \end{tablenotes}
  \end{threeparttable}
\end{table}

\subsubsection{Matrix A operands}
In the baseline approach, operands from matrix \textit{A} are fetched as scalars from the scalar register file and individually forwarded to the vector unit.
In contrast, the matrix algorithm retrieves multiple elements from \textit{A} in a tiled-vector manner, improving the access pattern and enabling the data reuse of the \textit{A} tile by means of the broadcast engine.
\subsubsection{Instruction count} 
In the baseline algorithm, each vector instruction is amortized over $vl$ operations.
With \gls{MX}, the total number of instructions fetched and decoded is lower, as each \texttt{mxfmacc} instruction is amortized over $m'n'k'$ operations, $m'k' = vl$, and $n' > 1$.
This boosts the \gls{SIMD} ratio, \ie the average number of operations per instruction.
\subsubsection{Tile window}
The matrix algorithm exploits the \textit{k} dimension to increase the size of the tile window when the dimensions \textit{M} and \textit{N} are limited.
This is especially beneficial as the \gls{SIMD} ratio is further improved by allowing each core to work on a larger output tile window in a multi-core environment when processing matrices with a limited $N$ dimension.
\subsubsection{Scalar-vector interactions} 
With the baseline algorithm, the scalar core must remain active to compute operand addresses and forward scalar operands to the \gls{VPU}. 
In contrast, \gls{MX} pushes the whole computation to the vector unit, freeing up the scalar core.
\subsubsection{Performance} 
The computing performance is significantly impacted by the number of data transfers between the memory and the \gls{VRF} and the related latency.
The \gls{MX}-ready \gls{VLSU} regularizes the memory accesses, which can reduce conflicts in both the interconnect and memory banks.
\subsubsection{Energy}
In the \gls{VPU}, the power consumption of the \gls{VRF} normally constitutes a non-negligible portion of the overall energy usage.
\gls{MX}'s inexpensive broadcast engine and tile buffers enhance data reuse for the tiled matrix \textit{A} and reduce the \gls{VRF} access by a \(\left(\frac{K}{k'}\right)\) factor.
Moreover, the reduced instruction count and more regular memory access pattern alleviate the pressure on the instruction and data memories, further improving the energy efficiency of the overall system.

\section{Experiment Setup and Results}
\subsection{Computing Clusters and Methodology}\label{sec:exp-setup}
We integrate the baseline and the \gls{MX}-ready versions of the Spatz \gls{VPU} into two floating-point-capable computing clusters: a \num{64}-bit constrained Dual-Core cluster for in-depth analysis of various tile and sub-tile configurations, and a \num{32}-bit large-scale 64-Core cluster for performance evaluation in a complex system.
\subsubsection{Dual-Core Cluster} 
The Dual-Core cluster is a \num{64}-bit shared-L1-memory cluster, implemented with \SI{128}{\kibi\byte} of \gls{TCDM} across \num{16} \gls{SRAM} banks.
This cluster features \num{2} Snitch cores, each controlling a Spatz instance equipped with \num{4} double-precision \glspl{FPU} and \SI{2}{\kibi\byte} \gls{VRF} each, supporting a vector length of \SI{512}{\bits}. The peak achievable performance is \SI{16}{DP-\flop\per\cycle}.
\subsubsection{64-Core MemPool Cluster} 
MemPool, a large-scale 32-bit shared-L1-memory cluster, scales up to \num{256} RISC-V cores and includes \SI{1}{\mebi\byte} of L1 \gls{TCDM}~\cite{MemPool2023_journal}.
The cluster is hierarchically organized into \num{4} groups, each containing \num{16} tiles. 
A fully connected logarithmic crossbar is employed between the cores and memories, achieving \gls{NUMA} with a maximum latency of \SI{5}{cycles}.
We equip each Spatz instance with \num{4} \num{32}-bit \glspl{FPU} and \SI{2}{\kibi\byte} of \glspl{VRF} each, supporting a vector length of \SI{512}{\bits}, and pair each instance with a scalar Snitch core to form a \gls{CC}.
This cluster configuration, labeled MemPool$_{\num{64}}$Spatz$_{\num{4}}$, consists of \num{64} \glspl{CC}, one for each tile, and achieves a peak performance of \SI{512}{SP-\flop\per\cycle}, as detailed further in~\cite{Spatz2022}.

We implement our designs in GlobalFoundries' \SI{12}{\nano\meter} LPPLUS FinFET technology through Synopsys Fusion Compiler 2022.03 for synthesis and \gls{PNR}.
We analyze the \gls{PPA} metrics of the \gls{MX}-ready clusters at the post-\gls{PNR} implementation stage and compare them to their respective non-\gls{MX} baseline architectures.  
We calculate power consumption using Synopsys' PrimeTime 2022.03 under typical operating conditions (TT/\SI{0.80}{\volt}/\SI{25}{\celsius}), with switching activities obtained from QuestaSim 2021.3 post-layout gate-level simulations and back-annotated parasitic information.
In the used \gls{MATMUL} kernels, all the input and output matrices are kept in the L1 memory and each core of the cluster calculates one portion of the output matrix.
The kernel executes in parallel across the entire cluster, partitioning the matrix equally among multiple cores.
At the end of each parallel task, the cores are synchronized to ensure consistent write-back of the results.

\subsection{Implementation Area and Frequency}\label{sec:exp-results-area-freq}
The logic area breakdown of the clusters is presented in~\Cref{tab:area-bdown}.
For the \gls{MX}-ready Dual-Core cluster, the main area increase originates from the \gls{VFU} (+5.3\%) due to the near-\gls{FPU} tile buffer and is followed by a slight increase in the \gls{VLSU} (+\SI{5.94}{kGE}), which is related to supporting matrix loads/stores.
The total area overhead of \gls{MX} is negligible, amounting to an increase of \SI{2.5}{\%}. 
The MemPool$_{\num{64}}$Spatz$_{\num{4}}$ cluster follows the same trend, resulting in a similar \SI{2.89}{\%} area overhead.
\gls{MX} does not affect the critical path of the two systems in analysis, which runs through Snitch to a \gls{TCDM} bank. Thus, the \gls{MX}-ready dual- and 64-core systems achieve 920 MHz and 720 MHz in the (SS/\SI{0.72}{\volt}/\SI{125}{\celsius}) corner, respectively, with no frequency degradation with respect to the baseline clusters.

\begin{table}[t]
  \caption{Logic Area Breakdown in 12-nm technology.}
  \label{tab:area-bdown}
  \centering
  \begin{threeparttable}
    \resizebox{\columnwidth}{!}{\begin{tabular}{lccrccr}
      \toprule & \multicolumn{3}{c}{Dual-Core Cluster{[}kGE{]}} & \multicolumn{3}{c}{64-Core Cluster{[}MGE{]}} \\
      \cmidrule(l){2-4} \cmidrule(l){5-7} 
      & \textbf{Baseline} & \textbf{MX} & \textbf{Overhead} & \textbf{Baseline} & \textbf{MX} & \textbf{Overhead}     \\
      \textbf{Snitch}  & \num{47.82}    & \num{48.01}    & +\num{0.40}\%   & \num{1.50}  & \num{1.47}  & -\num{2.04}\% \\
      \textbf{i-Cache} & \num{149.67}   & \num{149.56}   & -\num{0.07}\%   & \num{4.96}  & \num{4.95}  & -\num{0.20}\% \\
      \textbf{TCDM\tnotex{tnote:tcdm}}
                       & \num{1191.89}  & \num{1192.03}  & +\num{0.01}\%   & \num{20.46} & \num{20.48} & +\num{0.09}\% \\
      \textbf{VRF}     & \num{345.04}   & \num{348.87}   & +\num{1.11}\%   & \num{9.32}  & \num{9.32}  & \num{0.0}\%   \\
      \textbf{VFU}     & \num{1532.11}  & \num{1613.39}  & +\num{5.31}\%   & \num{12.91} & \num{13.97} & +\num{8.21}\% \\
      \textbf{VLSU}    & \num{111.66}   & \num{117.60}   & +\num{5.32}\%   & \num{2.54}  & \num{3.07}  & +\num{20.87}\%\\
      \textbf{Other}   & \num{570.63}   & \num{575.97}   & +\num{0.94}\%   & \num{7.28}  & \num{7.39}  & +\num{1.51}\% \\
      \midrule 
      \textbf{Total}   & \num{3948.82}  & \num{4045.43}  & +\num{2.45}\%   & \num{59.70} & \num{61.43} & +\num{2.89}\% \\
      \bottomrule
    \end{tabular}}
    \begin{tablenotes}
      \item\label{tnote:tcdm} Including Memory Banks and Interconnect Logic.
    \end{tablenotes}
  \end{threeparttable}
  \vspace{-1em}
\end{table}

\subsection{Performance, Power and Energy Efficiency}\label{sec:exp-results-perf-power}
\begin{table*}[t]
  \caption{The Summary of Kernel Information, Execution Performance and Energy Efficiency}
  \label{tab:dc-perf-eff}
  \centering
  \begin{threeparttable}
    \resizebox{\linewidth}{!}{\begin{tabular}{lrrrrrrrrrrr}
      \toprule
      \textbf{Config} & \textbf{\begin{tabular}[c]{@{}r@{}}Mtx Size\\ {[}M, N, K{]}\end{tabular}} & \textbf{\begin{tabular}[c]{@{}r@{}}Tile Size \\ {[}m, n, k{]}\end{tabular}} & \textbf{\begin{tabular}[c]{@{}r@{}}Sub-Tile Size \\ {[}m', n', k'{]}\end{tabular}} & \textbf{\begin{tabular}[c]{@{}r@{}}Mem-VRF \\ Transfers\end{tabular}} & \textbf{\begin{tabular}[c]{@{}r@{}}Arithmetic\\ Intensity\\ {[}FLOP/B{]}\end{tabular}} & \textbf{\begin{tabular}[c]{@{}r@{}}SIMD Ratio\\ {[}FLOP/vinsn{]}\end{tabular}} & \textbf{Utilization} & \textbf{\begin{tabular}[c]{@{}r@{}}Performance \\ @ss\_freq \\ {[}GFLOPS{]}\end{tabular}} & \textbf{\begin{tabular}[c]{@{}r@{}}Performance \\ @tt\_freq \\ {[}GFLOPS{]}\end{tabular}} & \textbf{\begin{tabular}[c]{@{}r@{}}Power \\ @tt\_freq \\ {[}W{]}\end{tabular}} & \textbf{\begin{tabular}[c]{@{}r@{}}En. Efficiency \\ @tt\_freq \\ {[}GFLOPS/W{]}\end{tabular}} \\
      \midrule
      \multicolumn{12}{c}{\textbf{Dual-Core Cluster}\tnotex{tnote:bold}\ \ \ \tnotex{tnote:dualcore}} \\
      \midrule
      Baseline & 64x64x64 & 8,16,1 & - & 53248 & 1.23 & 16.00 & 95.9\% & 14.13 & 15.34 & 0.21 & 71.49 \\
      \rowcolor{teal!20} Baseline & 64x64x64 & 4,32,1 & - & 77824 & 0.84 & 32.00 & \textbf{97.8\%} & 14.41 & 15.65 & 0.21 & \textbf{73.48} \\
      Baseline & 32x32x32 & 8,16,1 & - & 7168 & 1.14 & 16.00 & 90.0\% & 13.26 & 14.40 & 0.20 & 70.95 \\
      \rowcolor{teal!20} Baseline & 32x32x32 & 4,32,1 & - & 10240 & 0.80 & 32.00 & \textbf{93.3\%} & 13.75 & 14.93 & 0.20 & \textbf{72.87} \\
      \rowcolor{teal!20} Baseline & 16x16x16 & 8,16,1 & - & 1024 & 1.00 & 16.00 & \textbf{70.1\%} & 10.33 & 11.22 & 0.16 & \textbf{71.69} \\
      Baseline & 16x16x16 & 4,32,1 & - & 1408 & 0.73 & 32.00 & 64.7\% & 9.53 & 10.35 & 0.16 & 66.70 \\
      \midrule
      MX-ready & 64x64x64 & 4,8,4 & 4,4,4 & 102400 & 0.64 & 34.73 & 94.1\% & 13.86 & 15.06 & 0.21 & 72.91 \\
      MX-ready & 64x64x64 & 8,8,4 & 8,4,4 & 69632 & 0.94 & 63.22 & 95.6\% & 14.08 & 15.30 & 0.19 & 79.15 \\
      MX-ready & 64x64x64 & 4,16,4 & 4,4,4 & 86016 & 0.76 & 36.76 & 96.4\% & 14.20 & 15.42 & 0.21 & 75.19 \\
      \rowcolor{teal!20} MX-ready & 64x64x64 & 8,16,4 & 8,4,4 & 53248 & 1.23 & 66.59 & \textbf{97.2\%} & 14.32 & 15.55 & 0.19 & \textbf{81.49} \\
      MX-ready & 32x32x32 & 4,8,4 & 4,4,4 & 13312 & 0.62 & 34.29 & 88.4\% & 13.02 & 14.14 & 0.20 & 71.90 \\
      MX-ready & 32x32x32 & 8,8,4 & 8,4,4 & 9216 & 0.89 & 62.48 & 89.7\% & 13.22 & 14.35 & 0.18 & 77.68 \\
      MX-ready & 32x32x32 & 4,16,4 & 4,4,4 & 11264 & 0.73 & 36.21 & 92.7\% & 13.66 & 14.83 & 0.20 & 74.36 \\
      \rowcolor{teal!20} MX-ready & 32x32x32 & 8,16,4 & 8,4,4 & 7168 & 1.14 & 65.68 & \textbf{93.5\%} & 13.78 & 14.96 & 0.19 & \textbf{80.38} \\
      MX-ready & 16x16x16 & 4,8,4 & 4,4,4 & 1792 & 0.57 & 33.45 & 63.1\% & 9.30 & 10.10 & 0.15 & 67.45 \\
      MX-ready & 16x16x16 & 8,8,4 & 8,4,4 & 1280 & 0.80 & 61.09 & 66.1\% & 9.74 & 10.58 & 0.14 & 75.03 \\
      MX-ready & 16x16x16 & 4,16,4 & 4,4,4 & 1536 & 0.67 & 35.20 & 71.6\% & 10.55 & 11.46 & 0.16 & 72.03 \\
      \rowcolor{teal!20} MX-ready & 16x16x16 & 8,16,4 & 8,4,4 & 1024 & 1.00 & 64.00 & \textbf{70.3\%} & 10.36 & 11.25 & 0.15 & \textbf{75.41} \\ 
      \midrule
      \multicolumn{12}{c}{\textbf{64-Core Cluster}\tnotex{tnote:64core}}\\
      \midrule
      Baseline & 256x256x256 & 8,32,1  & - & 2686976 & 3.12 & 32 & 94.5\% & 372.26 & 439.94 & 1.57 & 279.86 \\
      Baseline & 128x128x128 & 8,32,1  & - & 344064  & 3.05 & 32  & 90.7\% & 357.34 & 422.31 & 1.57 & 268.64 \\
      Baseline & 64x64x64    & 8,8,1   & - & 69632   & 1.88 & 8   & 50.4\% & 198.57 & 234.68 & 1.20 & 194.91 \\
      \midrule
      MX-ready & 256x256x256 & 8,32,8  & 8,4,8 & 2686976 & 3.12 & 137.74 & 96.7\% & 380.74 & 449.97 & 1.46 & 307.35 \\
      MX-ready & 128x128x128 & 8,32,8  & 8,4,8 & 344064  & 3.05 & 136.23 & 95.8\% & 377.27 & 445.86 & 1.46 & 304.55 \\
      MX-ready & 64x64x64    & 8,8,8   & 8,4,8 & 69632   & 1.88 & 123.43 & 78.7\% & 309.99 & 366.35 & 1.50 & 244.24 \\
      \bottomrule 
    \end{tabular}}
    \begin{tablenotes}
      \scriptsize
      \item\label{tnote:bold} In bold, we highlight the best metrics for both the Baseline and MX-ready Dual-Core Cluster execution across various matrix sizes.
      \item\label{tnote:dualcore} Dual-Core Cluster: Double-Precision operations; $\mathrm{ss\_freq = 920 MHz; tt\_freq = 1 GHz}$.
      \item\label{tnote:64core} 64-Core Cluster: Single-Precision operations; $\mathrm{ss\_freq = 770 MHz; tt\_freq = 910 MHz}$.
    \end{tablenotes}
  \end{threeparttable}
  \vspace{-1em}
\end{table*}

\begin{figure}[htbp]
  \centering
  \begin{minipage}[h]{0.40\linewidth}
    \includegraphics[width=\linewidth]{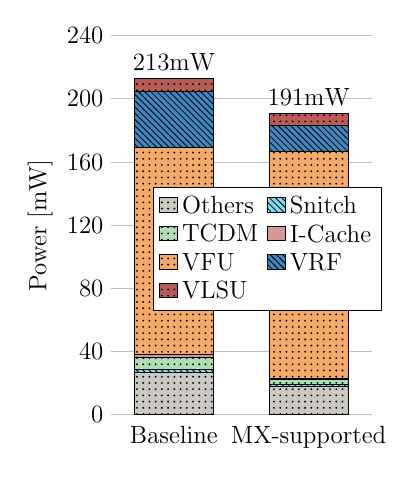}
  \end{minipage}
  \hspace{1em}
  \begin{minipage}[h]{0.40\linewidth}
    \includegraphics[width=\linewidth]{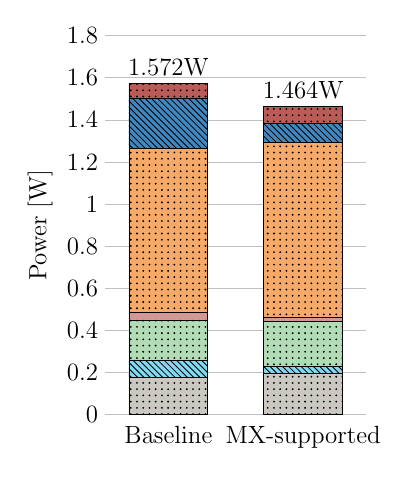}
  \end{minipage}
  \caption{Power breakdown for Dual-Core (Left) and 64-Core clusters (Right) executing MatMul. Dual-Core: at \textit{TT@1GHz}, executing non-\gls{MX} (4 vectors, length 32) and \gls{MX}-ready algorithms ($m'=8, n'=4, k'=4, B=4$). 64-Core: at \textit{TT@910MHz}, executing non-\gls{MX} (8 vectors, length 32) and \gls{MX}-ready algorithms ($m'=8, n'=4, k'=8, B=8$).}
  \label{fig:power_breakdown}
  \vspace{-1em}
\end{figure}

\subsubsection{Dual-Core Cluster}
The upper part of \Cref{tab:dc-perf-eff} summarizes the kernel information, execution performance, and energy efficiency for the Dual-Core cluster when executing \crdel{double-precision (\num{64}-bit)}\cradd{a \num{64}-bit} \gls{MATMUL} across various problem sizes and tile/sub-tile configurations, highlighting the rows where the kernel's tile and sub-tile configurations achieve the best energy efficiency.
The \gls{MX}-ready cluster with a sub-tile size of (\num{8}, \num{4}, \num{4}) achieves performance similar to the best-performing execution on the baseline cluster with efficiency gains by +\SI{10.9}{\%} (\numproduct{16x16x16}), +\SI{10.3}{\%} (\numproduct{32x32x32}), and +\SI{5.2}{\%} (\numproduct{64x64x64}).

We evaluate the baseline algorithm using two different output tile configurations with constant sizes.
For \crdel{a }small problem\cradd{s} \crdel{matrix size of}\cradd{(}\numproduct{16x16x16}\cradd{)}, the output tile size of \( (8,16,1) \) yields higher \gls{FPU} utilization.
As discussed in \Cref{sec:analysis}, although (\num{8}, \num{16}, \num{1}) has higher arithmetic intensity and fewer transfers between \gls{TCDM} and \gls{VRF} compared to (\num{4}, \num{32}, \num{1}), the latter configuration benefits from a $\num{2}\times$ increase in \gls{SIMD} ratio, leading to better performance for larger problem \crdel{matrix }sizes.
For the \gls{MX}-ready algorithm, the output tiles with larger \textit{$n$} and equal \textit{$n'$} sub-tile dimensions consistently yield better performance and energy efficiency. 
This improvement is attributed to their higher arithmetic intensity and average \gls{SIMD} ratio.
A similar trend is observed for the energy efficiency when increasing the \textit{$m'$} dimension of the sub-tile.
Due to the higher arithmetic intensity, the power \crdel{consumption }decreases when the output tile size changes from (\num{4}, \num{16}, \num{4}) to (\num{8}, \num{8}, \num{4}). 
However, the (\num{4}, \num{16}, \num{4}) configuration achieves higher performance thanks to more and shorter matrix result stores, which can be interleaved with computational instructions to hide latency.

The left part of \Cref{fig:power_breakdown} presents the power breakdown of the Dual-Core cluster's baseline and \gls{MX}-ready execution of a \numproduct{64x64x64} \gls{MATMUL}, with the most energy-efficient tile and sub-tile size in our benchmarks.
\gls{MX} reduces \gls{VRF} access for the B tile and intermediate result storage, leading to a \SI{53.5}{\%} reduction in \gls{VRF} power consumption.
Although the sub-tile buffer integration results in a slight \SI{9.4}{\%} power increase in \gls{VFU}, the overall \gls{VPU} power \crdel{consumption }decreases by \SI{4.1}{\%}.
We also observed a \cradd{power} decrease \crdel{in power consumption }across the rest of the cluster components, including the Snitch core, instruction caches, and \gls{TCDM}.
This reduction is attributed to the higher \gls{SIMD} ratio and tiled memory request pattern in \gls{MX}-ready execution, which eliminates the multiple requests for scalar operands generated by the Snitch core in the baseline\crdel{ execution}.
As a result, the total power savings for the \crdel{entire }Dual-Core cluster achieved through \gls{MX} amounts to \SI{10.4}{\%}.

\subsubsection{64-Core Cluster}
Our benchmark results for various problem sizes on MemPool$_{\num{64}}$Spatz$_{\num{4}}$ are presented in the bottom section of~\Cref{tab:dc-perf-eff}.
In such a large interconnected memory, contentions may occur when memory requests in the same tile access the same local bank or the same remote group in the same cycle.
This generates stalls of the \gls{VLSU} and increases the access latency.
Although such contentions could be mitigated by allocating data structures in a local tile's memory~\cite{Bertuletti2023}, this approach is hard to implement for \gls{MATMUL}, which inherently requires an extremely global data access pattern.

\gls{MX} regular memory accesses alleviate contention and improve \gls{VLSU} utilization by distributing vector element loads/stores across different banks and groups in a strided fashion, contrasting with the baseline\crdel{ cluster} where vector elements are fetched from continuous addresses within the same group by both scalar and vector core.
This is even more evident with small matrices\crdel{ like (\numproduct{64x64x64})}, where the initial vector load and final result store constitute a significant portion of the total runtime due to the inability to hide latency.
\gls{FPU} utilization increases from \num{50.4}\% to \num{78.7}\%, leading to a \num{56}\% improvement in cluster performance.

Despite a\crdel{n increase in} power consumption \cradd{increase} due to the higher \gls{FPU} utilization, the \gls{MX}-ready cluster achieves \SI{25}{\%} better energy efficiency.
Even though the baseline kernels already achieve near-peak utilization for matrix sizes of \numproduct{128x128x128} and \numproduct{256x256x256}, with the same arithmetic intensity, \gls{MX} still improves performance by \SI{5.6}{\%} and \SI{2.3}{\%}, with energy efficiency gains by \SI{13.4}{\%} and \SI{9.8}{\%}, respectively.
The right side of \Cref{fig:power_breakdown} presents the MemPool$_{\num{64}}$Spatz$_{\num{4}}$-related power breakdown comparison for a \numproduct{256x256x256} \gls{MATMUL}.
\gls{MX} reduces the \gls{VRF} power consumption by \SI{60}{\%}, thanks to fewer accesses achieved by buffering intermediate results.
The \gls{VFU} power \crdel{consumption }increases by only \SI{6}{\%}, which comes from the sub-tile buffer and higher \gls{FPU} utilization.
Overall, \gls{MX} leads to a \SI{6.9}{\%} cluster power reduction with near-peak \gls{FPU} utilization.

These analyses on both small-\crdel{scale} and large-scale vector \crdel{computing }clusters demonstrate that \gls{MX} significantly improves the energy efficiency by reducing the power consumption related to the \gls{VRF} accesses.
\gls{MX} also pushes the \gls{FPU} utilization closer to its peak with a negligible area overhead.
\cradd{A quantitative comparison of \gls{MX} against \cite{VERMA2022100742, thead-mmul-man} is hard since none of them presents area or power results, and the effective \gls{MATMUL} speed-up is unclear \cite{VERMA2022100742}.}

\section{Conclusion}\label{sec:conc}
In this paper, we presented \gls{MX}, an \gls{RVV}-based \gls{ISA} extension to support tiled matrix operations for energy-efficient \glspl{MATMUL}.
With an embedded-device-friendly and extremely low footprint overhead, \gls{MX} enhances the energy efficiency of \gls{MATMUL} by means of a small tile buffer near the \glspl{FPU}, which minimizes the\cradd{\ \gls{VRF}} accesses\crdel{ to the \gls{VRF}} by storing and reusing both input and output matrix tiles. 
Moreover, \gls{MX} reduces the number of instructions fetched by the scalar core, decreases the interaction between the scalar and vector cores, and regularizes the memory access pattern, further reducing power consumption.
We characterized \gls{MX} by implementing it on two multi-core clusters in a modern 12-nm technology node.
With less than \SI{3}{\%} area overhead and no impact on the operating frequency, \gls{MX} significantly boosts \gls{MATMUL}'s energy efficiency of a Dual-Core cluster by up to \SI{10.9}{\%}. \crdel{Furthermore, in}\cradd{In} a 64-Core cluster and \numproduct{64x64} matrices, performance and energy efficiency improve by \SI{56}{\%} and \SI{25}{\%}, respectively, further pushing the \gls{FPU} utilization toward the theoretical peak.

\section*{Acknowledgments}
This project has received funding from the ISOLDE project, No. 101112274, supported by the Chips Joint Undertaking of the  European Union’s Horizon Europe’s research and innovation program and its members Austria, Czechia, France, Germany, Italy, Romania, Spain, Sweden, Switzerland.

\bibliographystyle{IEEEtran}
\bibliography{refs}

\begin{thebibliography}{10}
\providecommand{\url}[1]{#1}
\csname url@samestyle\endcsname
\providecommand{\newblock}{\relax}
\providecommand{\bibinfo}[2]{#2}
\providecommand{\BIBentrySTDinterwordspacing}{\spaceskip=0pt\relax}
\providecommand{\BIBentryALTinterwordstretchfactor}{4}
\providecommand{\BIBentryALTinterwordspacing}{\spaceskip=\fontdimen2\font plus
\BIBentryALTinterwordstretchfactor\fontdimen3\font minus \fontdimen4\font\relax}
\providecommand{\BIBforeignlanguage}[2]{{%
\expandafter\ifx\csname l@#1\endcsname\relax
\typeout{** WARNING: IEEEtran.bst: No hyphenation pattern has been}%
\typeout{** loaded for the language `#1'. Using the pattern for}%
\typeout{** the default language instead.}%
\else
\language=\csname l@#1\endcsname
\fi
#2}}
\providecommand{\BIBdecl}{\relax}
\BIBdecl

\bibitem{PECCERILLO2022102561}
B.~Peccerillo, M.~Mannino, A.~Mondelli, and S.~Bartolini, ``A survey on hardware accelerators: Taxonomy, trends, challenges, and perspectives,'' \emph{Journal of Systems Architecture}, vol. 129, p. 102561, 2022.

\bibitem{AMIRI202083}
H.~Amiri and A.~Shahbahrami, ``Simd programming using {Intel} vector extensions,'' \emph{J. of Parallel and Distr. Comp.}, vol. 135, pp. 83--100, 2020.

\bibitem{9052677}
S.~Deng, H.~Zhao, W.~Fang, J.~Yin, S.~Dustdar, and A.~Y. Zomaya, ``Edge intelligence: The confluence of edge computing and artificial intelligence,'' \emph{IEEE Internet of Things Journal}, vol.~7, no.~8, pp. 7457--7469, 2020.

\bibitem{8358031}
N.~Jouppi, C.~Young, N.~Patil, and D.~Patterson, ``Motivation for and evaluation of the first tensor processing unit,'' \emph{IEEE Micro}, vol.~38, no.~3, pp. 10--19, 2018.

\bibitem{coral-bmark}
\BIBentryALTinterwordspacing
C.~AI, ``Edge {TPU} performance benchmarks,'' 2020. [Online]. Available: \url{https://coral.ai/docs/edgetpu/benchmarks}
\BIBentrySTDinterwordspacing

\bibitem{7738524}
Y.-H. Chen, T.~Krishna, J.~S. Emer, and V.~Sze, ``Eyeriss: An energy-efficient reconfigurable accelerator for deep convolutional neural networks,'' \emph{IEEE Journal of Solid-State Circuits}, vol.~52, no.~1, pp. 127--138, 2017.

\bibitem{VERMA2022100742}
V.~Verma, T.~{Tracy II}, and M.~R. Stan, ``{EXTREM-EDGE - EXtensions To RISC-V for Energy-efficient ML inference at the EDGE of IoT},'' \emph{Sust. Comp.: Informatics and Systems}, vol.~35, p. 100742, 2022.

\bibitem{thead-mmul-man}
\BIBentryALTinterwordspacing
{T-Head Semiconductor}, \emph{{RISC-V Matrix Multiplication Extension Specification}}, {T-Head Semiconductor}, 2023. [Online]. Available: \url{https://github.com/T-head-Semi/riscv-matrix-extension-spec/releases/tag/v0.3.0}
\BIBentrySTDinterwordspacing

\bibitem{Spatz2022}
M.~Cavalcante, D.~Wüthrich, M.~Perotti, S.~Riedel, and L.~Benini, ``Spatz: A compact vector processing unit for high-performance and energy-efficient shared-{L1} clusters,'' in \emph{Proc. of the 41st ICCAD}.\hskip 1em plus 0.5em minus 0.4em\relax San Diego, CA, USA: IEEE/ACM, Oct. 2022.

\bibitem{Kung1986}
H.~T. Kung, ``Memory requirements for balanced computer architectures,'' \emph{SIGARCH Comp. Arch. News}, vol.~14, no.~2, p. 49–54, May 1986.

\bibitem{billdallyHC}
B.~Dally, ``Hardware for deep learning,'' in \emph{Hot Chips}, Stanford, CA, USA, Aug. 2023.

\bibitem{Perotti2022}
M.~Perotti, M.~Cavalcante, N.~Wistoff, R.~Andri, L.~Cavigelli, and L.~Benini, ``{A `New {Ara}' for Vector Computing: an Open Source Highly Efficient {RISC-V} {V} 1.0 Vector Processor Design},'' in \emph{Proceedings of the 33rd IEEE Int. Conf. on ASAP}.\hskip 1em plus 0.5em minus 0.4em\relax Gothenburg, Sweden: IEEE, Jul. 2022.

\bibitem{MemPool2023_journal}
S.~Riedel, M.~Cavalcante, R.~Andri, and L.~Benini, ``{MemPool}: A scalable manycore architecture with a low-latency shared {L1} memory,'' \emph{IEEE Transactions on Computers}, 2023, early access.

\bibitem{Bertuletti2023}
M.~Bertuletti, Y.~Zhang, A.~Vanelli-Coralli, and L.~Benini, ``Efficient parallelization of {5G-PUSCH} on a scalable {RISC-V} many-core processor,'' in \emph{Proc. of the 2023 DATE Conf.}\hskip 1em plus 0.5em minus 0.4em\relax Antwerp, Belgium: {IEEE}, Mar. 2023.

\end{thebibliography}

\end{document}